\documentclass{article}
\usepackage{spconf,amsmath,graphicx,hyperref}
\usepackage{booktabs}
\usepackage{siunitx}
\usepackage{tabularx}
\usepackage{threeparttable}
\usepackage{multirow}
\usepackage{leadsheets}     
\usepackage{amsmath}
\usepackage{amssymb}

\newcommand{\fref}[1]{Fig.~\ref{#1}}

\title{
Rethinking music captioning with Music Metadata LLMs
}
\name{Irmak Bukey\thanks{*Work done while an intern at Adobe Research.}$^{\,\text{\sharp\:\flat*}}$ \quad Zhepei Wang$^{\,\text{\flat}}$ \quad Chris Donahue$^{\,\text{\sharp}}$ \quad Nicholas J.~Bryan$^{\,\text{\flat}}$}
\address{$^{\text{\sharp}}\,$ Carnegie Mellon University \quad \quad $^{\text{\flat}}\,$ Adobe Research
}
\begin{document}
\ninept
\maketitle

\begin{abstract}
Music captioning, or the task of generating a natural language description of music, is useful for 
both music understanding and controllable music generation. 
Training captioning models, however, typically requires high-quality music caption data which is scarce compared to metadata (e.g., genre, mood, etc.).
As a result, it is common to use large language models (LLMs) to synthesize captions from metadata to generate training data for captioning models, 
though this process imposes a fixed stylization and entangles factual information with natural language style.
As 
a more direct approach, we propose metadata-based captioning. We train a metadata prediction model to infer detailed music metadata from audio and then convert it into expressive captions via pre-trained LLMs at inference time.
Compared to a strong end-to-end baseline trained on LLM-generated captions derived from metadata, our method: 
(1) achieves comparable performance in less training time over end-to-end captioners,
(2) offers flexibility to easily change stylization post-training, enabling output captions to be tailored to specific stylistic and quality requirements, 
and 
(3) can be prompted with audio \textit{and} partial metadata to enable powerful metadata imputation or in-filling -- a common tasks for organizing music data.
\end{abstract}


\section{Introduction}
\label{sec:intro}

Large language models (LLMs) have proven to be a powerful foundation for music understanding, reasoning, captioning, and generation tasks for audio and music\cite{ltu, Qwen2-Audio, salmonn, flamingo3, llark, openmu, lpmusiccaps, chatmusician}. 
Multimodal LLMs (MLLMs), in particular, have made significant progress on music-related tasks by fine-tuning strong LLMs with paired audio and textual descriptions \cite{llark, openmu, chatmusician, mullama, musilingo, m2ugen}. Such approaches, however, are still limited by access to paired training data, due to the complexities of describing music via text, high licensing costs, and/or legal and/or ethical issues of web scraping.
Thus, accurate, informative, and diverse annotations or ``captions'' are critical.

Prior work has explored a range of strategies to 
synthesize or enhance music captioning datasets for training purposes. 
One approach is to use existing music caption datasets, often by combining several different sources \cite{musiclm, songdescriber}. However, these individual datasets are typically small in scale, limiting their usefulness for training robust models.
Another approach leverages LLMs to generate synthetic captions. 
Some methods augment existing captions or keyword-based descriptions \cite{Qwen2-Audio, flamingo3, m2ugen, laion_audio, augment} while others prompt LLMs to transform available music metadata and/or add so-called psuedo-labels via algorithmically extracting metadata from music information retrieval algorithms  (e.g. beats per minute) to create rich sentence-level captions or instruction-tuning datasets \cite{llark, lpmusiccaps, jamendomaxcaps}. 
This enables richer ground truth captions, but are still unsatisfactory as human and/or pseudo-label descriptions of music are typically imprecise and lack completeness.
Furthermore, many downstream tasks do not require captions, only rich metadata, so music captioning can become  
less flexible and lack control over language style, level of detail, and/or factual content once trained.

 \begin{figure*}[t]
        \centering 
        \includegraphics[width=\textwidth]{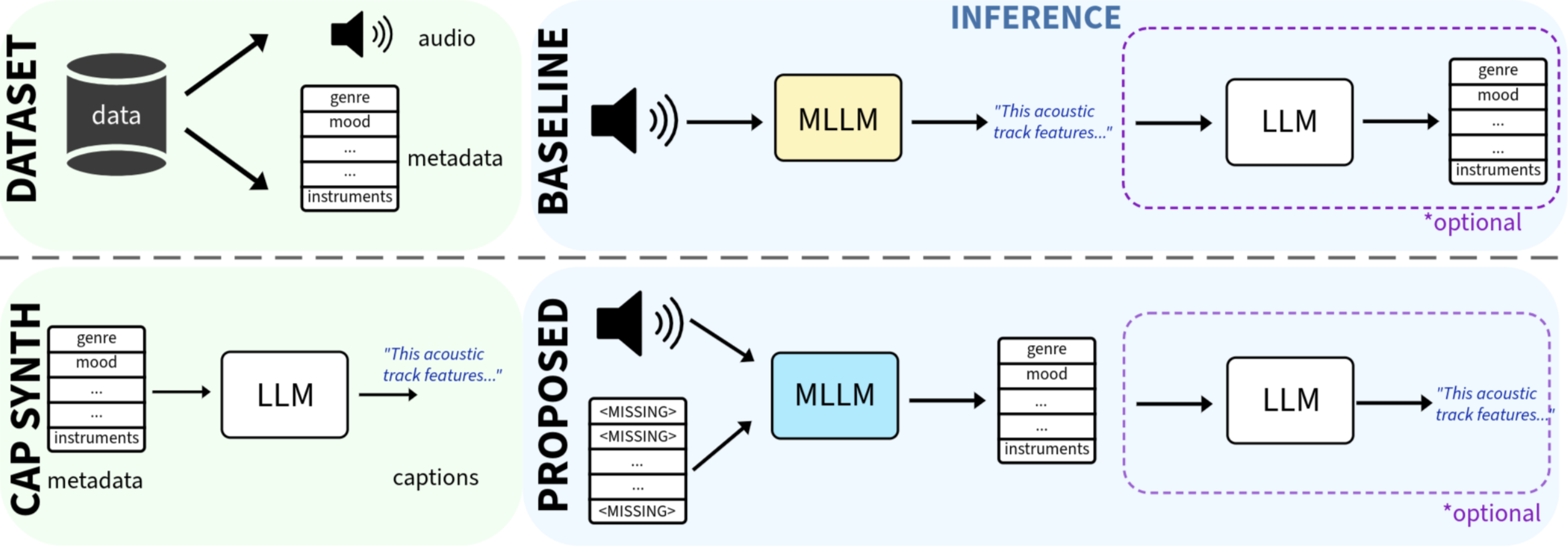} 
        \caption{(Top left) Dataset format. Our training dataset consists of paired music and metadata annotations. (Bottom left) Metadata-to-caption synthesis. A text-based LLM is used to convert music metadata into natural language descriptions. (Top right) Inference pipeline for typical music captioning methods where a caption MLLM is trained to generate captions. Optionally, a text-based LLM can be used to extract medatata from generated captions. (Bottom right) Inference pipeline for the proposed metadata MLLM that predicts all or missing metadata fields. Optionally, a text-based LLM can be used to convert the predicted metadata into captions.}
        \label{fig:headline} 
    \end{figure*}

In this work, we propose music \textit{metadata} LLMs -- an alternative approach to music captioning that is easily adaptable to different natural language styles as shown in~\fref{fig:headline} (bottom right). Specifically, we instruction-tune a pretrained LLM on a metadata prediction task to learn to infer structured music metadata with an open-set vocabulary directly from audio. 
Our predicted metadata is then transformed into captions using 
the same pretrained LLM via a text prompt command.  
Furthermore, we train our metadata LLM with zero or more input metadata fields in addition to audio inputs, so we can use our model to impute missing metadata fields of an existing music database.
For evaluation, we compare our approach to a strong end-to-end baseline trained on LLM-generated captions derived from metadata, test the impact of post-hoc stylization, and show the impact of including partial metadata in addition to audio input.
Overall, we find that our proposed method offers competitive performance, faster training, greater flexibility in caption style and quality, and the ability to leverage partial metadata for improved metadata prediction—unlike end-to-end captioners trained on fixed captions. We summarize our contributions: 
\begin{enumerate}
    \item We propose an audio-to-metadata-to-caption method for music, with comparable performance and faster training relative to audio-to-caption methods. 
    \item We demonstrate that post-hoc metadata-to-caption conversion can improve overall captioning performance and offers increased flexibility. 
    \item 
    We show that our audio-to-metadata model effectively leverages partial metadata for metadata imputation, resulting in improved prediction performance.
\end{enumerate}

\section{Relevant Work}

Early works on music captioning include MusiCNN+LSTM \cite{musicnn} which formulates captions as concatenation of music tags which lack expressiveness of natural language, and MusCaps \cite{muscaps}. More recent works including Qwen2-Audio \cite{Qwen2-Audio}, SALMONN \cite{salmonn}, Audio Flamingo 3 \cite{flamingo3}, LLark \cite{llark}, Lp-MusicCaps \cite{lpmusiccaps}, and ViML \cite{viml}
construct training data for music understanding by leveraging LLMs to rephrase existing metadata, human-labeled captions, and/or pseudo-labels from content analysis algorithms into an instruction-following format. While this approach enables scalable supervision through natural language and advances general music understanding and reasoning tasks, the resulting captions often lack musical details such as instrumentation, key, and tempo that are desirable for downstream applications with fine-grained semantic control. Retrieval-based works like TTMR++ \cite{ttmr} also use metadata and LLMs to enrich training captions, but do not train to generate detailed descriptions.

Meanwhile, in the image captioning domain, structured metadata representations have been explored as intermediaries for image captioning pipelines. 
For instance, an image-to-scene-graph-to-caption pipeline~\cite{sgae,sgae_distill} can lead to richer captions, which in turn can lead to improved image generation quality~\cite{betker2023improving}.
This approach is increasingly relevant as improvements to language models decrease the burden of the second stage~\cite{tfsgc}. 
However, to the best of our knowledge, analogous approaches have not been explored for music captioning.

\section{Method}
\label{sec:format}

Our method generates captions via two-stages. First, we predict structured metadata from audio and optional partial input metadata. Second, we convert the predicted metadata into music captions using an LLM-based conversion stage for stylization.

\subsection{Metadata prediction model}

We instruction fine-tune modern decoder-only pretrained text-only LLMs using audio-metadata pairs to perform metadata prediction from audio. Since we start with a pretrained text-only LLM, 
we adopt a two-stage finetuning procedure to adapt it for audio inputs and structured output generation.

In the first stage, we imbue the text-only LLM with multimodal audio understanding capabilities. To this end, we first encode audio inputs into a sequence of discrete tokens using a quantized audio encoder~\cite{casebeer2025} and then map the audio tokens to reserved text tokens in the LLM.
We then adapt the pre-trained LLM on self-supervised audio and language continuation tasks simultaneously, resulting in a pretrained audio-text MLLM.

In the second stage, we instruction fine-tune the 
audio MLLM model on a metadata prediction task. Our training examples consist of metadata in JSON format and audio pairs as illustrated in~\fref{fig:headline} (top left). 
The objective, as expressed in detail in our instruction prompt, is to predict metadata fields in a structured format while keeping the provided fields unchanged. Our setup enables the model to perform metadata 
imputation, 
effectively learning to infer a complete set of musical attributes from both the audio and 
(optionally) 
partial contextual information. 
At inference time, we provide audio and an empty metadata dictionary as input to our model, prompting it to generate the entire set of metadata fields in a structured manner. 

\subsection{Metadata-to-caption conversion}
Given predicted metadata, we convert it into captions using the original text-only LLM. We do this with a carefully designed instruction to guide the model to generate an expressive, coherent caption that reflects the predicted metadata fields while avoiding hallucinations of information not present in the predicted metadata output. Furthermore, we can easily adapt the metadata-to-caption prompt post-hoc to customize for a given problem domain or target stylization by incorporating in-context learning examples \cite{gpt3, dong2022survey}. This is in contrast to the existing approach of synthesizing captions from metadata \emph{before} training.
We explore different variations for our post-hoc caption conversion in our experiments below to evaluate their impact on caption quality and controllability.

\subsection{Metadata imputation}
Beyond our standard use case of predicting music metadata and music captions, we also address the task of music metadata \emph{imputation}---predicting complete metadata from audio and an incomplete subset of metadata. 
Metadata imputation is a common real-world task for music data organization and useful for both re-training new captioners and/or metadata models. 
To support imputation, 
we artificially mask available metadata fields in the training data, and prompt our model to generate the complete metadata from the audio tokens and the masked metadata. 
We note that this is a very natural task to add to our core approach, but more difficult to do with captioning models since there is no easy way to on-the-fly modify pre-generated captions to omit specific metadata fields.

\section{Experiments}
\label{sec:pagestyle}

\subsection{Experimental setup}
We compare our metadata captioning method against baseline captioners trained end-to-end on audio-caption pairs. All models are evaluated on 1) metadata prediction and 2) caption generation.

\textbf{Datasets.} We train the model with an internal licensed instrumental music dataset containing approximately 25k hours of music with corresponding metadata annotations with fields such as genre, mood, keywords, tempo, key, energy, and instruments. Note that the metadata is often incomplete, as $23\%$ of the music entries miss one or more fields. The audio training examples are randomly selected 10-second audio chunks from these tracks. For evaluation of metadata prediction, we use a held-out subset of this internal dataset with 5k tracks. For evaluation of caption generation, we use music-caption pairs from the non-vocal subset of the MusicCaps dataset, containing 2,185 tracks \cite{musiclm, sao}, and from the Song Describer dataset, containing 446 tracks \cite{songdescriber}.

\textbf{Baselines.} We train two end-to-end music captioners as baselines for comparison. Using the same metadata dataset as our method, we prompt Gemma3-12B-it to generate captions by distilling metadata in the styles of MusicCaps and SongDescriber captions, respectively. This produces two sets of (audio, synthetic caption)  pairs, each aligned with a specific captioning style. We then instruction fine-tune our audio-adapted model on these pairs, resulting in two baseline captioners: one trained to produce MusicCaps-style captions, and the other Song Describer-style captions.

\textbf{Training configurations.} For all models, we instruction fine-tune Gemma3-1B-it~\cite{gemma3}, a decoder-only text-only LLM, which we imbue with multimodal audio capabilities using the method outlined above. For metadata-to-caption and caption-to-metadata conversions, we prompt the original Gemma3-1B-it model, without any additional finetuning. The proposed metadata model and the baseline captioners are all trained on 4 NVIDIA A100 GPUs
with early stopping, resulting in checkpoints at 161,000 iterations for the metadata model and 347,600 iterations for both captioners.
For our audio autoencoder, we use an architecture based on DAC \cite{dac},
but with an encoder with more aggressive temporal downsampling~\cite{rebottleneck} that consists of 32 channels with a frame rate of 21 Hz, which is further quantized with a codebook size of 1024~\cite{casebeer2025}.

\subsection{Evaluation tasks}
We evaluate our audio-to-metadata model and audio-to-caption baseline on both metadata and captioning, using LLMs to adapt each to the task they were not trained on.

\textbf{Metadata evaluation.} For the proposed method, we prompt the model to generate full structured metadata and evaluate the predictions for semantic fields including genre, mood, instruments, and keywords. For baseline outputs, we extract metadata from generated captions by prompting an LLM with field-specific questions (e.g. \textit{``What is the genre of this track as inferred from the caption?"}). This allows us to construct structured metadata for the baseline captioner outputs and prepare them for metadata evaluation. To evaluate, each predicted and ground truth field is converted into a sentence using fixed templates (e.g. the instruments prediction \textit{[electric guitar, drums, bass]} would be converted into the sentence \textit{``This track features the instruments electric guitar, drums, bass."}).
We compute similarity between sentence-BERT (SBERT) embeddings \cite{sbert} between predictions and ground truth to handle varying-length inputs, synonyms and open-ended vocabulary.

\textbf{Caption evaluation.} For caption evaluation, we compute SBERT similarity between predicted and reference captions, both for our pipeline and baselines. 
To more thoroughly analyze the effect of style, we explore two complementary setups involving cross evaluation:
(1) cross-style prompting for the proposed metadata-to-caption pipeline, where our model generates metadata for MusicCaps audios but use the Song Describer-style prompts for metadata-to-caption conversion (and vice versa), and (2) cross-dataset evaluation for the baseline captioners, where the captioner trained on MusicCaps-style captions is evaluated on the Song Describer dataset (and vice versa). 

\subsection{Style and prompting variations}
An advantage of our method is its flexibility in adapting to new caption styles without retraining. To explore this, we experiment with
additional metadata-to-caption prompts and evaluate impact on caption quality. For comparison, we also apply post-hoc stylization prompts to the outputs of baseline captioners to assess whether similar stylistic control can be achieved by modifying captions directly. In particular, we focus on variants of in-context learning or providing rich exemplars in an LLM prompt. We experiment with using a fixed example set, randomly sampled example set, and metadata tags as part of the example set for in-context learning.  

For these experiments, we use stylistic similarity metrics to measure the impact of different style prompts on captioning performance. We combine three complementary metrics: BM25 \cite{bm25}, a bag-of-words retrieval-based measure that captures lexical overlap and content relevance; length similarity, computed with a smoothed exponential function of the caption length difference; and part-of-speech (POS) histogram similarity, comparing the syntactic structure of predicted and reference captions. We also report the average of all semantic and stylistic metrics.


Lastly, we set up experiments to evaluate our model’s metadata prediction performance when it is prompted with partial metadata at inference time, simulating settings where some metadata information is available and can be used for filling in the missing fields.

\section{Results and Analysis}
\label{sec:typestyle}


\begin{table}[t]
\setlength\tabcolsep{4pt}
\centering
\begin{minipage}{\columnwidth}
\begin{threeparttable}
\caption{Metadata prediction performance of Metadata and MusicCaps-style (MC) and Song Describer-style (SD) captioner models, evaluated using the SBERT similarity. Higher is better.}
\label{tab:md_eval}
\begin{tabular*}{\columnwidth}{@{\extracolsep{\fill}}lccccc}
\toprule
\textbf{Model} & \textbf{Genre} & \textbf{Mood} & \textbf{Instr.} & \textbf{Kwrds.} & \textbf{Avg} \\
\midrule
MC Captioner    & 0.556 & 0.673 & \textbf{0.677} & 0.614 & 0.630 \\
SD Captioner    & \textbf{0.562} & 0.687 & 0.676 & \textbf{0.618} & \textbf{0.636} \\
Metadata (Ours) & 0.548 & \textbf{0.711} & 0.675 & 0.566 & 0.625 \\
\bottomrule
\end{tabular*}
\end{threeparttable}
\end{minipage}
\end{table}

\begin{table}[t]
\centering
\caption{
Caption evaluation results using SBERT similarity for matched and cross-style/dataset captioning setups, evaluated on MusicCaps (MC) and Song Describer (SD) datasets. Higher is better.
}
\label{tab:caption_eval}
\begin{tabularx}{\columnwidth}{l l >{\centering\arraybackslash}X >{\centering\arraybackslash}X >{\centering\arraybackslash}X}
\toprule
\textbf{Style} & \textbf{Model} & \textbf{MC} & \textbf{SD} & \textbf{Avg} \\
\midrule
\multirow{2}{*}{Matched} 
& Captioner & \textbf{0.478} & 0.468 & \textbf{0.407} \\
& Metadata (Ours) & 0.443 & 0.454 & 0.392 \\
\midrule
\multirow{2}{*}{Cross} 
& Captioner & 0.441 & \textbf{0.469} & 0.405 \\
& Metadata (Ours)  & 0.439 & 0.462 & 0.395 \\
\bottomrule
\end{tabularx}
\end{table}

\begin{table*}[t]
\small
\centering
\caption{Impact of different style prompts on captioning performance. ``Captioner" results represent two baseline captioners under the matched style condition. For the proposed work, we experiment with different style prompts during the metadata-to-caption stage. The ``fixed 1-shot'' uses a fixed instead of randomly sampled in-context learning example (reference style caption). ``metadata 1-shot'' includes metadata tags in the example. For baseline captioner models, we use equivalent style prompts to post-hoc edit the predicted captions. Higher is better.}
\begin{tabularx}{\linewidth}{l|cc|c|cc|c|cc|c|cc|c|c}
\toprule
\multirow{2}{*}{\textbf{Method}} & \multicolumn{3}{c|}{\textbf{SBERT-Sim} } & \multicolumn{3}{c|}{\textbf{BM25}} & \multicolumn{3}{c|}{\textbf{Length}} & \multicolumn{3}{c|}{\textbf{POS} } & \multirow{2}{*}{\textbf{Avg}}  \\
\cmidrule(r){2-4} \cmidrule(lr){5-7} \cmidrule(lr){8-10} \cmidrule(lr){11-13} 
 & \textbf{MC} & \textbf{SD} & \textbf{Avg} & \textbf{MC} & \textbf{SD} & \textbf{Avg} & \textbf{MC} & \textbf{SD} & \textbf{Avg} & \textbf{MC} & \textbf{SD} & \textbf{Avg} &  \\
\midrule
Captioner & 0.478 & 0.468 & 0.473 & 0.177 & 0.104 & 0.141 & 0.293 & 0.122 & 0.208 & 0.805 & \textbf{0.724} & \textbf{0.765} &0.396\\
+shorter prompt & 0.477 & 0.466 & 0.472 & 0.174 & 0.101 & 0.138 & 0.293 & 0.138 & 0.216 & 0.803 & 0.722 & 0.763 &0.397\\
+fixed 1-shot & 0.476 & 0.467 & 0.472 & 0.174 & 0.104 & 0.139 & 0.296 & 0.138 & 0.217 & 0.804 & \textbf{0.724} & 0.764 &0.398\\
\midrule
Metadata & 0.443 & 0.454 & 0.449 & 0.201 & \textbf{0.111} & 0.156 & 0.205 & 0.165 & 0.185 & 0.768 & 0.702 & 0.735 &0.381\\
+shorter prompt & 0.450 & 0.463 & 0.457 & 0.175 & 0.089 & 0.132 & 0.252 & 0.234 & 0.243 & 0.774 & 0.707 & 0.741 &0.393\\
+fixed 1-shot & 0.475 & \textbf{0.475} & 0.475 & 0.190 & 0.059 & 0.125 & \textbf{0.319} & 0.412 & 0.366 & 0.787 & 0.694 & 0.741 &0.426\\
+metadata 1-shot & \textbf{0.495} & 0.471 & \textbf{0.483} & \textbf{0.317} & 0.045 & \textbf{0.181} & 0.284 & \textbf{0.454} & \textbf{0.369} & \textbf{0.810} & 0.656 & 0.733 &\textbf{0.442}\\
\bottomrule
\end{tabularx}
\label{tab:style_experiments}
\end{table*}


\begin{table}[t]
\centering
\caption{SBERT scores for partial metadata completion. \% refers to proportion of other fields available (predicted field always masked). Ours-0\% is equivalent to first row of Table~\ref{tab:md_eval}. Higher is better.}
\begin{tabular}{lccccc}
\toprule
\textbf{Model} & \textbf{\%} & \small{\textbf{Genre}} & \small{\textbf{Mood}} & \small{\textbf{Instr.}} & \small{\textbf{Kwrds.}} \\
\midrule
Gemma3-1b & 50\%	&0.504	&0.666	&0.657	&0.543\\
\midrule
\multirow{5}{*}{Ours} & 0\%    & 0.548 & 0.711 & 0.675 & 0.566\\
 & 25\%  & 0.638 & 0.743 &   0.754	& 0.618\\
 & 50\%  & 0.679 & 0.765 &  0.78	& 0.645 \\
 & 75\%  & 0.715 & 0.789 &  0.807	& 0.671 \\
 & 100\%  & \textbf{0.731} & \textbf{0.798} &  \textbf{0.817}	& \textbf{0.686} \\
\bottomrule
\end{tabular}
\label{tab:partial_md_exps}
\end{table}

First, we empirically show that the proposed music metadata LLMs obtain comparable performance to the baseline audio-to-caption methods on both metadata prediction and caption generation tasks with significantly lower training costs. Table~\ref{tab:md_eval} reports metadata prediction results using the SBERT similarity metric for the fields genre, mood, instruments, and keywords. Our proposed method achieves performance comparable to baseline captioners on average, with better mood predictions while worse at keywords. We observe similar performance comparison for captioning evaluation in Table~\ref{tab:caption_eval}. We attribute the competitive performance of the baseline captioners in part to their training on captions generated from the same metadata used in our pipeline, resulting in closely aligned training examples between the two compared approaches. Moreover, a key consideration is the inductive bias of the base LLM, which is pretrained on natural language data more similar to that of the baseline captioners, rather than the more constrained and structured JSON data our model is trained on. This inductive bias likely benefits the baseline captioners. Nonetheless, our model achieves similar performance on both metadata and caption generation while being trained for only $46.3\%$ of the GPU hours compared to baselines.

From Table~\ref{tab:caption_eval}, we also observe in the ``Cross'' row that using MusicCaps-style reference captions when captioning Song Describer audio does not lead to a performance drop, whereas the reverse setting does. This is likely because MusicCaps captions tend to be more detailed and descriptive than the generally shorter and less specific Song Describer captions, therefore still maintaining decent performance across different datasets.

Next, we demonstrate our proposed audio-to-metadata pipeline offers flexibility for post-hoc stylization and further improves captioning performance. In Table~\ref{tab:style_experiments}, we present the impact of prompt engineering on our proposed method.
As we refine the metadata-to-caption conversion prompts, we observe consistent improvements in output quality and style, improving captioning performance by more than $20\%$ without any retraining. In contrast, applying similar prompts for post-hoc editing of baseline captioner outputs does not yield comparable gains and results in only negligible changes, highlighting a strength of our method.

Finally, we demonstrate another key advantage of our model: the ability to leverage partial metadata to improve metadata prediction performance and enable metadata imputation. In Table~\ref{tab:partial_md_exps}, we evaluate our model's metadata prediction performance by progressively increasing the amount of partial metadata available at inference time. For benchmarking, we also include an LLM baseline (Gemma3-1b-it) and assess its metadata prediction performance when prompted with $50\%$ of the available fields. Our results show that prompting our model with partial metadata significantly boosts prediction performance, achieving an average improvement of $21\%$ and up to $33\%$ on individual fields.  For downstream tasks like using metadata and/or captions to condition music generation models, an ability to impute high-quality metadata to complete database entries is a significant benefit. Similar partial prompting and targeted prediction is not feasible with end-to-end captioners.

\section{Conclusion}
\label{sec:majhead}
In this paper, we present a flexible and effective approach to music captioning by training an audio-to-metadata prediction model and converting the rich, structured metadata predictions into captions using a pre-trained LLM at inference time. The ability to convert metadata to captions post-hoc offers increased flexibility for controlling caption style while also improving overall captioning performance. Additionally, our model’s metadata prediction performance is enhanced when prompted with partial metadata, allowing for effective metadata imputation—a capability not supported by end-to-end captioning pipelines, 
but highly valuable for music data organization. Our results demonstrate the effectiveness and flexibility of this metadata-driven pipeline while also highlighting several key advantages over traditional captioning methods.

For future work, we believe there are many exciting directions. We are excited about the idea of training with dynamic, changing structured metadata schemas, length-controlled fields to allow adapting from single to multi-label fields at inference-time.


\vfill\pagebreak




\clearpage
\bibliographystyle{IEEEbib}
\bibliography{strings,refs}

\end{document}